\documentclass[sigconf]{acmart}

\AtBeginDocument{%
  \providecommand\BibTeX{{%
    \normalfont B\kern-0.5em{\scshape i\kern-0.25em b}\kern-0.8em\TeX}}}

\setcopyright{acmcopyright}
\copyrightyear{2023}
\acmYear{2023}
\acmDOI{XXXXXXX.XXXXXXX}

\usepackage{hyperref}
\usepackage{subfig}
\usepackage{bm}
\usepackage{amsthm}
\usepackage{algorithm}
\usepackage{algorithmic}
\usepackage{pifont}
\usepackage{multirow}
\usepackage{array}
\usepackage{booktabs}

\usepackage{bbm}
\usepackage{colortbl}  
\usepackage{xcolor}
\usepackage{array} 

\renewcommand{\paragraph}[1]{\vspace{1ex}\noindent\textbf{#1}.}
\begin{document}

\title{DiffusionCom: Structure-Aware Multimodal Diffusion Model for Multimodal Knowledge Graph Completion}

\author{Wei Huang}
\affiliation{
  \institution{Beijing University of Posts and Telecommunications}
  \city{Beijing}
  \country{China}
}
\author{Meiyu Liang}
\affiliation{
  \institution{Beijing University of Posts and Telecommunications}
  \city{Beijing}
  \country{China}
}
\author{Peining Li}
\affiliation{
  \institution{Beijing University of Posts and Telecommunications}
  \city{Beijing}
  \country{China}
}
\author{Xu Hou}
\affiliation{
  \institution{Beijing University of Posts and Telecommunications}
  \city{Beijing}
  \country{China}
}
\author{Yawen Li}
\affiliation{
  \institution{Beijing University of Posts and Telecommunications}
  \city{Beijing}
  \country{China}
}
\author{Junping Du}
\affiliation{
  \institution{Beijing University of Posts and Telecommunications}
  \city{Beijing}
  \country{China}
}
\author{Zhe Xue}
\affiliation{
  \institution{Beijing University of Posts and Telecommunications}
  \city{Beijing}
  \country{China}
}
\author{Zeli Guan}
\affiliation{
  \institution{Beijing University of Posts and Telecommunications}
  \city{Beijing}
  \country{China}
}

\renewcommand{\shortauthors}{}

\begin{abstract}
Multimodal knowledge graphs (MKGs) have been widely applied in various downstream tasks, including recommendation systems, information retrieval, and visual question answering. However, existing knowledge graphs remain significantly incomplete, prompting the rapid development of multimodal knowledge graph completion (MKGC) methods. Most current MKGC approaches are predominantly based on discriminative models that maximize conditional likelihood. These approaches struggle to efficiently capture the complex connections in real-world knowledge graphs, thereby limiting their overall performance. To address this issue, we propose a structure-aware multimodal \textbf{Diffusion} model for multimodal knowledge graph \textbf{Com}pletion (\textbf{DiffusionCom}). DiffusionCom innovatively approaches the problem from the perspective of generative models, modeling the association between the $(head, relation)$ pair and candidate tail entities as their joint probability distribution $p((head, relation), (tail))$, and framing the MKGC task as a process of gradually generating the joint probability distribution from noise. Furthermore, to fully leverage the structural information in MKGs, we propose Structure-MKGformer, an adaptive and structure-aware multimodal knowledge representation learning method, as the encoder for DiffusionCom. Structure-MKGformer captures rich structural information through a multimodal graph attention network (MGAT) and adaptively fuses it with entity representations, thereby enhancing the structural awareness of these representations. This design effectively addresses the limitations of existing MKGC methods, particularly those based on multimodal pre-trained models, in utilizing structural information. DiffusionCom is trained using both generative and discriminative losses for the generator, while the feature extractor is optimized exclusively with discriminative loss. This dual approach allows DiffusionCom to harness the strengths of both generative and discriminative models. Extensive experiments on the FB15k-237-IMG and WN18-IMG datasets demonstrate that DiffusionCom outperforms state-of-the-art models. Notably, on FB15k-237-IMG, DiffusionCom achieves a 38.2\% relative improvement in Hits@1 compared to previous leading methods.
\end{abstract}



\keywords{Multimodal Knowledge Graph Completion, Multimodal Graph Attention Networks, Conditional Denoising Diffusion, Relation Extraction}

\maketitle

\section{INTRODUCTION}
\begin{figure}[t]
\centering
\includegraphics[width=0.48\textwidth]{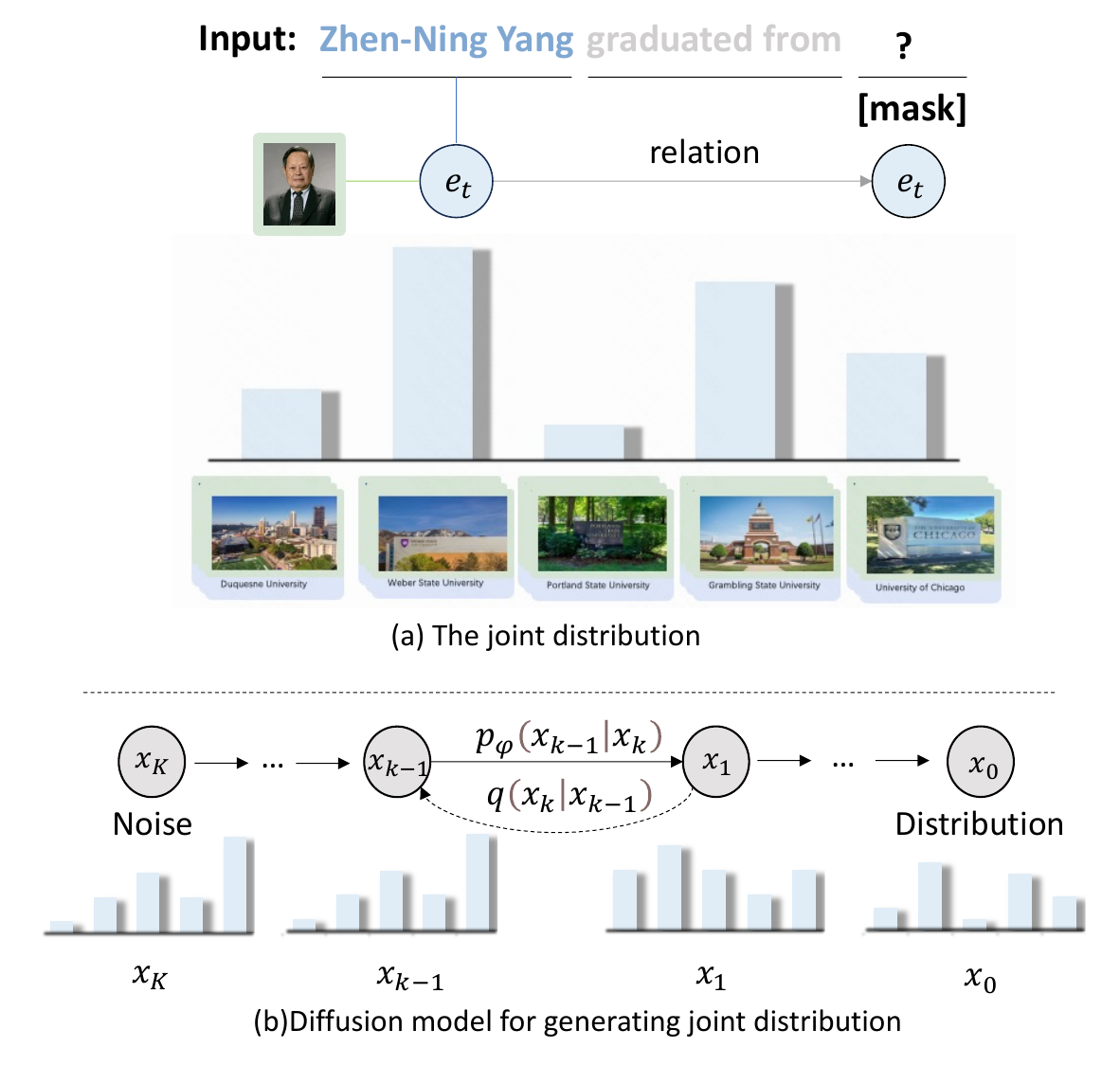}
\caption[DiffusionCom Model]{DiffusionCom for multimodal knowledge graphs completion. (a) We propose to model the correlation between the $(head, relation)$ and the candidate tail entities as their joint probability. (b) Diffusion models have demonstrated strong generative capabilities across various fields. Leveraging their coarse-to-fine generative characteristics, we employ diffusion models to generate joint probabilities.}
\label{introduction}
\end{figure}

Knowledge Graphs (KGs) represent real-world data in the form of factual triples $(head, relation, tail)$, demonstrating broad application potential in various fields such as intelligent recommendation systems \cite{zhang2021alicg,huang2018improving}, information retrieval \cite{2017Explicit,2020Biomedical,dietz2018utilizing}, and visual question answering \cite{hao2017end}. Multimodal Knowledge Graphs (MKGs), by integrating data from different modalities like text and images, provide richer and more accurate knowledge representations for a wide range of tasks \cite{FB15K-237}. However, MKGs still face the challenge of knowledge incompleteness due to the limited availability of multimodal corpora and the complexity of emerging entities and relations. To address this issue, the task of multimodal knowledge graph completion has been proposed \cite{ComplEX,AdaMF-MAT,MoSE,zhang2024making,zhang2024native,chen2024continualmultimodalknowledgegraph}. The goal of this task is to enhance entity embeddings using multimodal data, to uncover missing information in the graph, thereby completing the structure and content of the multimodal knowledge graph.

Pre-trained Transformer architecture \cite{vaswani2017attention} has achieved remarkable success across various domains \cite{liang2022semantic,xiong2023confidence,xiong2024temporal,xiong2025uniattnreducinginferencecosts}. Inspired by these advancements, multimodal pre-trained Transformer (MPT) models have been developed for multimodal knowledge graph completion task, such as KG-BERT \cite{kg-bert}, MKGformer \cite{MKGformer}, MPKGAC \cite{wang2023mpkgac} and SGMPT \cite{SGMPT}. Unlike other multimodal data, multimodal knowledge graphs (MKGs) typically encompass three types of information, textual information $\mathcal{M}_t$(e.g., textual descriptions), visual information $\mathcal{M}_v$(e.g., images), and graph-structured information $\mathcal{G}$. In early MPT-based approaches, the structure of the knowledge graph is primarily utilized to match and retrieve images and textual descriptions of the same entity \cite{kg-bert,MKGformer,wang2023mpkgac}. Although SGMPT \cite{SGMPT} attempts to incorporate structural information into the MPT framework, its approach is relatively simplistic, considering only the structural information of a single triplet. As a result, existing MPT methods fail to fully leverage the rich structural information of knowledge graphs, such as the relations between entities and the topological characteristics of graphs. This limitation hinders the model performance when handling more complex reasoning tasks.

Moreover, these approaches inherently belong to discriminative models. From a probabilistic perspective, discriminative models can only learn the conditional probability distribution, i.e., $p((tail)|(head, relation))$. This approach allows the model to focus solely on the marginal distribution between the conditions and the target, emphasizing the extraction of known relational patterns from existing data. Since it does not require in-depth modeling of the underlying data distribution, discriminative models are typically limited to handling single relations or simple associative features, making it difficult to effectively capture the complex interactions between multiple latent factors \cite{bernardo2007generative,liang2022gmmseggaussianmixturebased}. For example, in biomedical knowledge graphs, the relations between entities often exhibit high complexity and diversity. The associations between genes and diseases are usually not simple linear or one-to-one relations, but are influenced by various biological processes, environmental factors, epigenetics, and more. In such scenarios, models that rely on conditional probability distributions tend to overlook the interactions between these complex factors, resulting in a significant decline in predictive performance in intricate settings.

From the above, two significant challenges arise in the task of multimodal knowledge graph completion (MKGC). Challenge 1: Existing models that optimize conditional probability distribution struggle to capture the underlying multimodal data distribution, limiting their capacity to account for all relations in complex, real-world multimodal knowledge graphs. Challenge 2: Most models based on multimodal pre-trained Transformer (MPT) primarily focus on textual ($\mathcal{M}_t$) and visual ($\mathcal{M}_v$) semantic information, but do not fully utilizing the structural information embedded in the multimodal knowledge graph ($\mathcal{G}$).

To address the two challenges mentioned above, we propose a structure-aware multimodal \textbf{diffusion} model for multimodal knowledge graph \textbf{com}pletion (\textbf{DiffusionCom}), which models the completion task as a process of progressively generating joint distributions from noise and achieves a deep perceptual fusion of semantic and structural information. For Challenge 1, motivated by the remarkable advancements of diffusion models in various discriminative tasks \cite{jin2023diffusionret,wu2020diffnet,li2023diffurec,wu2020diffnet}, we set out to consider the multimodal knowledge graph completion (MKGC) task from a generative perspective and proposed a multimodal diffusion model. As illustrated in Figure \ref{introduction}, Given a set of $(head, relation)$ pair and target tail entities, the multimodal diffusion model is employed to gradually generate their joint probability distribution from noise. To enhance the performance of the generative model, we optimize the proposed method from both generation and discrimination perspectives. During the training phase, the generator is optimized not only with a conventional generative loss but also with a discriminative loss, while the feature encoder is trained solely through the discriminative loss. For Challenge 2, we propose Structure-MKGformer as the encoder for DiffusionCom. Structure-MKGformer employs the Multimodal Graph Attention Network (MGAT) \cite{GAT} to reason over the graph and effectively capture underlying fine-grained structural relationships. An adaptive weighted fusion strategy is then applied to integrate structural information.

Our contribution can be summarised as follows:
\begin{itemize}
\item We propose a novel structure-aware multimodal \textbf{diffusion} model for multimodal knowledge graph \textbf{com}pletion \textbf{(DiffusionCom)}. DiffusionCom integrates semantic and structural information through multimodal fusion and adaptive structural learning. Then frame the multimodal knowledge graph completion task as a joint distribution generation problem and employ a Denoising Diffusion Probabilistic Models (DDPM) to directly generate the joint probability distribution, offering a new perspective on this task. To the best of our knowledge, this is the first attempt to explore the potential of diffusion models for multimodal knowledge graph completion tasks.

\item We propose a novel conditional denoiser for multimodal diffusion models, which integrates a constrained multimodal conditioning mechanism to learn the reverse diffusion process and generate joint distributions from noisy data.
\item We propose Structure-MKGformer, an adaptive, structure-aware multimodal knowledge representation learning method based on the Multimodal Graph Attention Network (MGAT) \cite{GAT}, as the encoder for DiffusionCom. Structure-MKGformer fully leverages the rich semantic and structural features present in multimodal knowledge graphs, thoroughly explores implicit fine-grained structural relationships between entities, and employs an adaptive weighted fusion strategy to effectively integrate structural information.
\item Extensive experiments on the FB15k-237-IMG and WN18-IMG datasets demonstrate the superiority of DiffusionCom in the multimodal knowledge graph completion task. Especially on FB15k-237-IMG, DiffusionCom achieves a 38.2\% relative improvement in the Hits@1 metric compared to the current state-of-the-art methods.
\end{itemize}

\section{RELATED WORK}
\subsection{Multimodal Knowledge Graph Completion}
Multimodal Knowledge Graph Completion (MKGC) aims to infer missing entities by integrating information from different modalities such as text and images. There are two main architectures in MKGC:
\textbf{Non-Transformer-based architecture} in MKGC is built upon classic models such as TransE \cite{TransE}, which embeds relations using translations in a low-dimensional space, and DistMult \cite{DistMult}, which employs tensor decomposition to represent triples as embeddings of head, relation, and tail entities. ComplEx \cite{ComplEX} extends these methods by using complex-valued embeddings. IKRL \cite{IKRL} combines text and image features with structural embeddings. TransAE \cite{TransAE} takes this a step further by using a multimodal auto-encoder with TransE, incorporating visual and textual information into final entity representations. RSME \cite{RSME} applies a relation-sensitive filter to prioritize relevant visual contexts, adjusting its input based on task relevance. MoSE \cite{MoSE}, based on existing methods, uses modality-specific embeddings and ensemble inference, learning separate representations for each modality to avoid interference. LAFA \cite{LAFA} introduces a link-aware fusion and aggregation approach.

With the rise of transformer-based models, new approaches have emerged that leverage the power of pre-trained Transformers for even more advanced multimodal knowledge graph completion.

\textbf{Transformer-based architecture} has become the dominant paradigm in multimodal tasks due to their superior performance \cite{CLIP,li2022blip,li2024llava}. Early attempts to directly apply general multimodal pre-trained Transformer models (such as VisualBERT \cite{visualbert}, ViLBERT \cite{Vilbert}) to MKGC task. To better support MKGC task, the VBKGC \cite{vbkgc} model uses pre-trained Transformers to encode multimodal features and designs a specialized multimodal scoring function. DRAGON \cite{DRAGON} employs a self-supervised learning approach to pre-train text and knowledge graphs. MKGformer \cite{MKGformer} proposes a hybrid transformer framework with a multi-level fusion mechanism. Based on this, SGMPT \cite{SGMPT} further incorporates structural information to improve the model performance. HRGAT \cite{HRGAT} enhances reasoning capabilities by aggregating multimodal features through a hypernode graph. Recently, the MRE \cite{MRE} model proposes an end-to-end framework that achieves zero-shot relation learning in multimodal knowledge graph completion. MyGO \cite{MyGO} significantly improves the reasoning ability for missing knowledge through fine-grained modality handling, cross-modal entity encoding, and contrastive learning. AdaMF-MAT \cite{AdaMF-MAT} assigns adaptive weights to each modality and generates adversarial samples, thus enhancing underutilized modality information and improving the accuracy and efficiency of multimodal knowledge graph completion. 

However, all these methods are based on discriminative models, inferring missing relations by maximizing conditional likelihood. In complex relational multimodal knowledge graph completion task, it is difficult to capture all connections. In contrast, the proposed DiffusionCom leverages denoising diffusion probabilistic model (DDPM) \cite{diffusion(ddpm)} to learn the underlying data distribution and reformulates the entity prediction task as modeling the joint probability distribution.
\subsection{Diffusion Models}
Diffusion models \cite{sohl2015deep,diffusion(ddpm),dhariwal2021diffusion} are a class of generative models inspired by thermodynamic principles. Specifically, the process involves gradually injecting Gaussian noise into the data following a pre-defined noise schedule until reaching a final time step $K$. Then, a neural network is trained to progressively denoise the data, reversing the process to generate the target data. In recent years, diffusion models have achieved remarkable success in generative tasks, such as image generation \cite{ho2022classifier,wang2022zero,ho2022cascaded,hatamizadeh2025diffit}, natural language generation \cite{gulrajani2024likelihood,lin2023text,lovelace2024latent}, and audio generation \cite{huang2023make}. Additionally, some studies have explored the potential of applying diffusion models to discriminative tasks, including image segmentation \cite{wolleb2022diffusion,rahman2023ambiguous,wu2024medsegdiff}, multimodal recommendation systems \cite{wu2020diffnet,li2023diffurec}, and detection \cite{chen2023diffusiondet}. Recently, FDM \cite{long2024fact} has initially explored the application of diffusion models in knowledge graph completion, but has not yet extended it to multimodal scenarios. 

By efficiently modeling complex data distributions, diffusion models have significantly improved performance across various tasks. However, despite their impressive achievements in numerous fields, the application of diffusion models to the task of multimodal knowledge graph completion remains underexplored. This study fills this gap by modeling the correlation between $(head, relation)$ and candidate tail entities as their joint probability, and employing diffusion model to progressively generate the joint probability distribution from noise. To the best of our knowledge, we are the first to apply diffusion models to the task of multimodal knowledge graph completion.
\section{METHODOLOGY}
\begin{figure*}[t]
\centering
\includegraphics[width=0.98\textwidth]{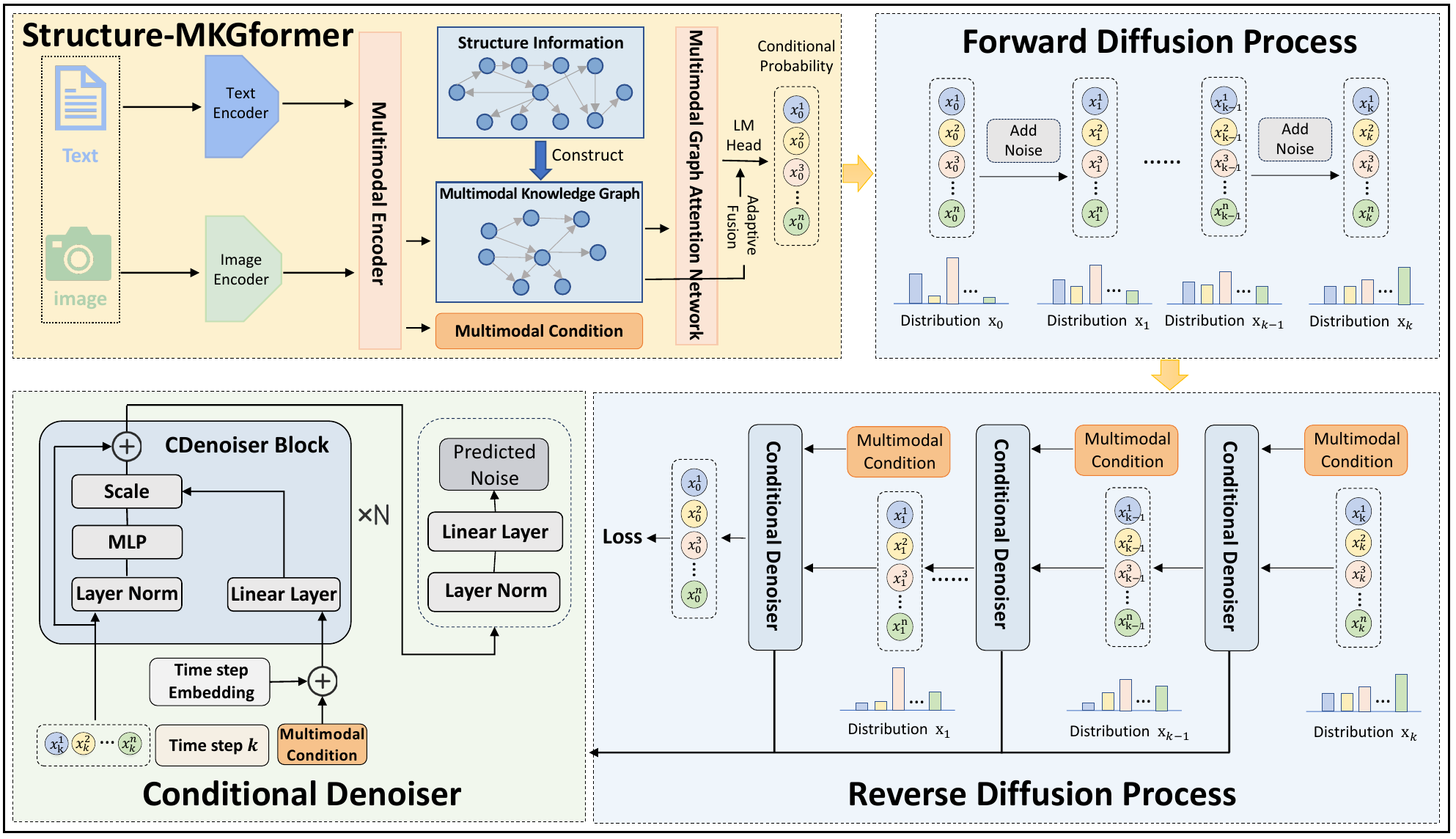}
\caption{Framework of the proposed DiffusionCom method.}
\label{pipeline}
\end{figure*} 

In this section, we will provide a detailed explanation of the formal description and implementation process of the model. First, we will introduce the specific problem definition of the MKGC task. Then, we will elaborate step by step on the overall process of DiffusionCom and the implementation details of each module. Finally, we will discuss the training strategy of the model and the design of the corresponding loss function.

\subsection{Problem Formulation}
A multimodal knowledge graph (MKG) is defined as a directed graph $M K G=(\mathcal{E}, \mathcal{R}, \mathcal{G}, \mathcal{A}_{\mathcal{M}})$, where $\mathcal{E}$ and $\mathcal{R}$ represent the sets of entities and relations, respectively. $\mathcal{G}=\{(e_{h}, r_{h, t}, e_{t}) \mid e_{h}, e_{t} \in \mathcal{E}, r_{h, t} \in \mathcal{R} \}$ is the set of fact triples. $\mathcal{A}_{\mathcal{M}}$ denotes the set of multimodal attributes associated with each entity, comprising two modalities: textual descriptions $\mathcal{M}_t$ and visual descriptions $\mathcal{M}_v$. The goal of multimodal knowledge graph completion (MKGC) is to predict the missing entities in $\mathcal{G}$, i.e., given an incomplete fact $(h, r, ?)$, we aim to predict the missing tail entity $t$. Noticing that the problem $(?, r, t)$ is the same, this paper only discusses $(h, r, ?)$.

\subsection{DiffusionCom Framework}
The overall framework of DiffusionCom is illustrated in Figure \ref{pipeline}. It consists of two main components: Structure-MKGformer and conditional denoiser (CDenoiser). Specifically, we propose the Structure-MKGformer, which employs a multimodal graph attention network to capture the structural relations between entities, thereby enhancing the structural awareness of entity representations. Structure-MKGformer integrates both visual and textual modality information to generate a multimodal condition embedding and simultaneously produce a conditional probability distribution. This distribution is progressively noised to form Gaussian noise, which is subsequently denoised step by step through CDenoiser to generate the joint probability distribution.

\subsection{Structure-MKGformer}
As illustrated the Figure \ref{pipeline}, Structure-MKGformer builds the text encoder, image encoder, and multimodal encoder following the settings of MKGformer \cite{MKGformer}. Specifically, the number of layers in the text encoder, image encoder, and multimodal information encoder are denoted as $L_t$, $L_v$, and $L_m$, respectively. Notably, $L_{BERT} = L_t + L_m$ and $L_{ViT} = L_v + L_m$. Below, we briefly introduce the three key components: text encoder, image encoder, and multimodal encoder (see \cite{MKGformer} for further details).

\textbf{Text Encoder}. The text encoder ${f}_t(\cdot )$ consists of the first $L_t$ layers of BERT\cite{devlin2018bert} and is designed to capture essential syntactic and lexical information. It takes the tokenized text descriptions $\mathcal{M}_t$ as input and produces the corresponding textual features, denoted as $\mathcal{H}_t={f}_t(\mathcal{M}_t)$.

\textbf{Image Encoder}. The image encoder ${f}_v(\cdot )$ consists of the first $L_v$ layers of ViT\cite{dosovitskiy2020image}. Its purpose is to capture fundamental visual features from images. It takes the image $\mathcal{M}_v$ as input and produces the corresponding visual features, denoted as $\mathcal{H}_v={f}_v(\mathcal{M}_v)$.

\textbf{Multimodal Encoder}. The multimodal encoder ${f}_m(\cdot )$ aims to model the multimodal features of entities through multi-level fusion, using the final $L_m$ layers of both ViT and BERT. It takes the representations learned by the previous encoders as input and outputs multimodal representations, denoted as $\mathcal{H}_m={f}_m(\mathcal{H}_t,\mathcal{H}_v)$.

\begin{figure}[t]
\centering
\includegraphics[width=0.45\textwidth]{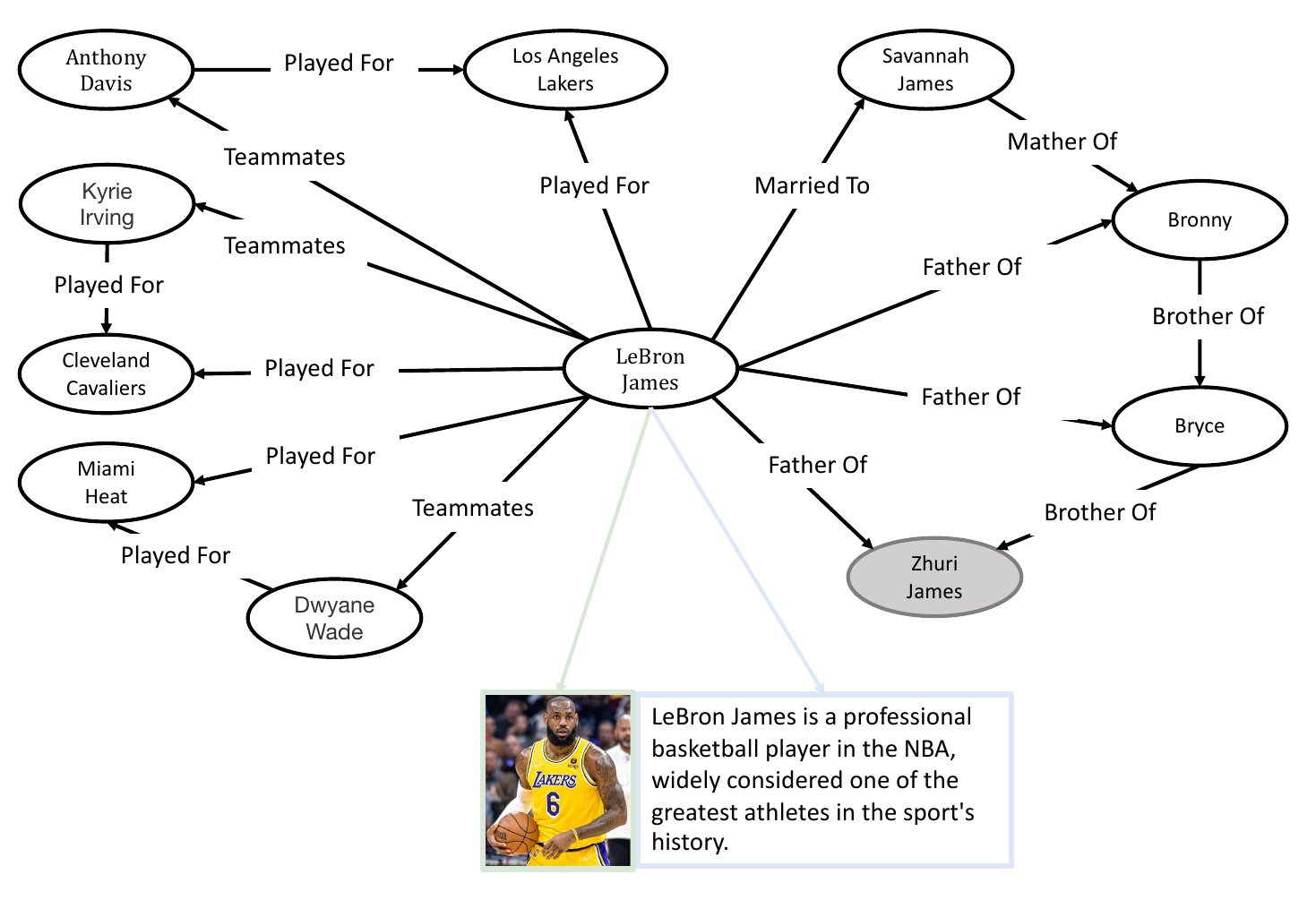}
\caption{An example in the knowledge graph.}
\label{SMKG}
\end{figure}

To fully utilize the structured information in multimodal knowledge graphs (MKGs), we integrate a mask token that encapsulates multimodal information with the structural information to create a multimodal knowledge graph. We then apply reasoning through the Multimodal Graph Attention Network (MGAT) to enable deep learning of implicit and fine-grained structural relationships. Figure \ref{SMKG} shows an example of a multimodal knowledge graph (MKG) using LeBron James. Various relationships like "father of," "teammates," and "played for" are linked to him. The left side focuses on his role as a basketball player, while the right highlights his role as a father. These different perspectives, however, can introduce noise. For example, when predicting Zhuri James as the missing entity in "LeBron James father of Zhuri James," relationships from his basketball career may not help and could even interfere with the model prediction. To address this, we have optimized the representation of structural information in Structure-MKGformer. Specifically, during tail entity prediction, a local subgraph is generated based on the head entity and relation, filtering out irrelevant structural information. This strategy effectively reduces the model's burden of processing overly complex relational networks and minimizes noise interference.

The entire Multimodal Graph Attention Network (MGAT) can be represented as $Z_{mask} = MGAT(H_{mask}, {\mathcal{G}}')$, where $H_{mask}$ denotes the embedding of the Mask token that incorporates the text and image modality information, and ${\mathcal{G}}'=subgraph(G,h,r)$ denotes a subgraph of the multimodal knowledge graph extracted from the structural information $\mathcal{G}$, based on the entity $h$ and the relation $r$. The reasoning results obtained by the MGAT over the MKG are denoted as $Z_{mask}$. Then we perform an adaptive weight fusion of $H_{mask}$ and $Z_{mask}$, the formula is $Z^{'}_{mask}=H_{mask}*\lambda+Z_{mask}*(1-\lambda)$, where $\lambda$ is a learnable parameter. Subsequently, a conditional probability distribution $x_0=p(tail\mid(head, relation))$ is formed through a Language Model Head (LM Head).

\subsection{Diffusion Process}
Our method reformulates the multimodal knowledge graph completion task from a generative modeling perspective. Inspired by the remarkable success of diffusion models across various domains \cite{diffusion(ddpm),hatamizadeh2025diffit}, we adopt them as the generator in our framework. Specifically, for a given $(head, relation)$ pair and $N$ candidate tail entities, our objective is to generate a distribution $x^{1: N}=\left\{x^{i}\right\}_{i=1}^{N}$ starting from Gaussian noise $\mathcal{N} (0, I)$. Unlike prior works that primarily focus on optimizing the posterior probability $p(t\mid h,r;\theta_m)$, our method aims to establish a joint probability distribution 
\begin{equation}
x^{1: N}=p((h,r), t \mid \phi)=f_{\phi}((h,r), t, \mathcal{N}(0, \mathrm{I}))
\end{equation}
where $\phi$ denotes the parameters of the generator. 

As illustrated in Figure \ref{pipeline}, the overall diffusion process can be summarized into two primary stages: the forward diffusion process and the reverse diffusion process with conditional denoising. Specifically, during the forward diffusion stage, the model progressively adds Gaussian noise to the input data according to a predefined noise scheduling scheme, gradually transforming it into Gaussian distribution by timestep $K$. Subsequently, in the reverse diffusion stage, DiffusionCom learns to model the Markov transition from Gaussian distribution to the distribution of plausible facts in the vector space through the Conditional Denoiser (CDenoiser). To further improve the quality of the reverse diffusion process, explicit conditional constraints are incorporated into the CDenoiser. These constraints integrate input conditions and enable the effective learning of diverse connection patterns during the reverse diffusion process. The following sections will elaborate in detail on the implementation and design of these two core stages.

\textbf{Forward Diffusion Process}. In the forward diffusion process $q(x_k\mid x_{k-1})$, noise sampled from a Gaussian distribution is added to the probability distribution $x_0$. This process progressively maps the factual embedding $x_0$ into pure noise by iteratively applying noise at each timestep $T_k=i$ until the diffusion step reaches $T_k=K$. Each transition $x_{T_{k-1}} \to  x_{T_{k}}$ is parametrized by:
\begin{equation}
q\left(x_{T_{k}} \mid x_{T_{k}-1}\right)=\mathcal{N}\left(x_{T_{k}} ; \sqrt{1-\beta_{T_{k}}} x_{T_{k}-1}, \beta_{T_{k}} I\right)
\end{equation}
where$\{\beta_{T_{k}}\}_{T_k=1}^{K}$are forward process variances. This parametrization of the forward process contains no trainable parameters.

\begin{algorithm}[t]
\caption{The First stage}
\label{The first stage}
\begin{algorithmic}[1]
\STATE Input: Textual descriptions $\mathcal{M}_t$, Visual descriptions $\mathcal{M}_v$, Structural information $\mathcal{G}$, Batch size, Number of epochs;
\STATE Parameters: Text encoder ${f}_t$, Image encoder ${f}_v$, Multimodal information encoder ${f}_m$, LM head $LM_{head}$, Multimodal Graph Attention Network $MGAT$;
\STATE OutPut: Structure-MKGformer;
\FOR{epoch in 1 to Number of Epochs}
    \FOR{ batch of data}
        \STATE $\mathcal{H}_t={f}_t(\mathcal{M}_t)$;
        \STATE $\mathcal{H}_v={f}_v(\mathcal{M}_v)$;
        \STATE $\mathcal{H}_m={f}_m(\mathcal{H}_t,\mathcal{H}_v)$;
        \STATE $\mathcal{H}_{mask}\gets gather(\mathcal{H}_m,Mask_{id})$;
        \STATE $x_0 = LM_{head}(MGAT(\mathcal{H}_{mask}, \mathcal{G}))$;
        \STATE Take the gradient descent step on;
        \STATE $\mathcal{L}_{D} = y \cdot \log (Sigmoid(x_0))+(1-y) \cdot \log (1-Sigmoid(x_0))$;
        \STATE Backpropagate and update Structure-MKGformer;
    \ENDFOR
\ENDFOR
\RETURN Structure-MKGformer;
\end{algorithmic}
\end{algorithm}

\begin{algorithm}[t]
\caption{The Second stage}
\label{The second stage}
\begin{algorithmic}[1]
\STATE Input: Textual descriptions $\mathcal{M}_t$, Visual descriptions $\mathcal{M}_v$, Structural information $\mathcal{G}$, Batch size, Number of epochs;
\STATE Parameters: Conditional Denoiser;
\STATE OutPut: Conditional Denoiser;
\FOR{epoch in 1 to Number of Epochs}
    \FOR{ batch of data}
        \STATE Calculate $x_0$ by Structure-MKGformer.;
        \STATE $T_{k} \sim \operatorname{Uniform}(\{1, \cdots, K\})$;
        \STATE $\boldsymbol{\epsilon} \sim \mathcal{N}(\mathbf{0}, \mathbf{I})$;
        \STATE $\hat{x}_{T_{k}}=\sqrt{\bar{\alpha}_{T_{k}}} x_0+\sqrt{1-\bar{\alpha}_{T_{k}}} \boldsymbol{\epsilon}$;
        \STATE $\operatorname{Denoise}\left(\hat{x}_0\right) = 
            \frac{1}{\sqrt{\bar{\alpha}_{T_{k}}}} \hat{x}_{T_{k}} - $
            \begin{align*}
                & \sqrt{\frac{1}{\bar{\alpha}_{T_{k}}}-1} \epsilon_{\theta}\left(\hat{x}_{T_{k}}, T_{k}, \hat{x}_{c}\right)
            \end{align*}
        \STATE Take the gradient descent step on;
        \STATE $\mathcal{L}_{CD}=\mathcal{L}_{G}+\mathcal{L}_{D}$;
        \STATE Backpropagate and update Conditional Denoiser;
    \ENDFOR
\ENDFOR
\RETURN Conditional Denoiser;
\end{algorithmic}
\end{algorithm}

\textbf{Reverse Diffusion Process}. In the reverse diffusion process, DiffusionCom defines a conditional reverse diffusion process, denoted as $p\left(\hat{x}_{T_{k}-1} \mid \hat{x}_{T_{k}}, \hat{x}^{c}\right)$, where $\hat{x}^{c}$ represents the Multimodal Condition embedding from Structure-MKGformer and $\hat{x}_K=x_K$. 

This process conditions on $\hat{x}^{c}$ and iteratively denoises the initial Gaussian noise to gradually approximate the target probability distribution. The transition between adjacent latent variables can be expressed as:
\begin{equation}
\begin{aligned}
p_{\theta}\left(\hat{x}_{T_{k}-1} \mid \hat{x}_{T_{k}}, \hat{x}^{c}\right)=\mathcal{N}\left(\hat{x}_{T_{k}-1}; \mu_{\theta}\left(\hat{x}_{T_{k}}, T_{k}, \hat{x}^{c}\right), \sigma_{T_{k}}^{2} I\right)
\end{aligned}
\end{equation}
Here $\sigma_{T_k}$ represents the constant variance as defined in\cite{diffusion(ddpm)}, and $\mu_{\theta}$ denotes the mean of the Gaussian distribution computed by the denoiser, where $\mu$ represents the parameters of the neural network. According to\cite{diffusion(ddpm)}, we can reparameterize the mean to enable the neural network to learn the noise introduced by the timestep $T_k$. Consequently, the expression for $\mu_{\theta}$ can be reparameterized as follows:
\begin{equation}
\mu_{\theta}(\hat{x}_{T_{k}}, T_{k}, \hat{x}^{c})=\frac{1}{\sqrt{\alpha_{T_{k}}}}(\hat{x}_{T_{k}}-\frac{\beta_{T_{k}}}{\sqrt{1-\bar{\alpha}_{T_{k}}}} \epsilon_{\theta}\left(\hat{x}_{T_{k}}^{\tau}, T_{k}, \hat{x}^{c}\right))
\end{equation}
where $\{\beta_{T_k}\}_{T_k=1}^T$ denote the variance in the forward process, $\alpha_{T_{k}}=1-\beta_{T_{k}}$ and $\bar{\alpha}_{T_{k}}=\prod_{s-1}^{T_{k}} \alpha_{s}$. $\epsilon_{\theta}\left(\hat{x}_{T_{k}}, T_{k}, \hat{x}^{c}\right)$ is the neural network to predict the added noise conditioned on known condition embeddings at time step $T_k$. This neural network is referred to as the Conditional Denoiser (CDenoiser), which will be explained in detail in the next section.

\textbf{Conditional Denoiser}. In the reverse process of Denoising Diffusion Probabilistic Models (DDPM), designing an appropriate denoising model is crucial. Currently, most existing denoising models in DDPM are primarily designed for image or text data, while the data format of knowledge graphs differs. Knowledge graphs are typically represented as triples $(h, r, t)$, with relatively simple data structures and less significant long-range dependencies. In response to this, we propose a simple and efficient Conditional Denoiser (CDenoiser) based on multi-layer perceptron (MLP) architecture, specifically tailored for processing knowledge graphs, as opposed to the commonly used transformer backbone. Formally, the architecture of CDenoiser can be described as follows:
\begin{equation}
\hat{x}^{\boldsymbol{ct}}=\text { LinearLayer }\left(\hat{x}^{c}+\hat{x}_{T_k}^T\right)
\end{equation}
\begin{equation}
E=\text { CDenoiserBlock }\left(\hat{x}_{T_{k}}, \hat{x}^{\boldsymbol{ct}}\right)
\end{equation}
\begin{equation}
\epsilon=\text { LinearLayer }(\operatorname{Layernorm}(E))
\end{equation}
where $\hat{x}^{\boldsymbol{ct}}$ is the final condition embedding calculated by $\hat{x}_c$ and timestep embedding $\hat{x}_{T_k}^T$ at step $T_k$, $\hat{x}_{T_{k}}$ is the noised fact embedding at step $T_k$. $E$ is intermediate feature calculated by the CDenoiser block, and $\epsilon$ is the noise predicted by CDenoiser. Next, we will introduce the CDenoiser block.

\textbf{CDenoiser Block}. Inspired by the success of transformer encoders \cite{transformer} in the field of graph data\cite{LAFA}, the CDenoiser Block adopts a similar architecture. It is composed of MLP layers, with Layernorm (LN) applied before each layer and residual connections\cite{2016Deep} used at the end of each sublayer. As mentioned earlier, the form of facts $(h,r,t)$ is simple and short, with less evident long-range dependencies. CDenoiser replaces the multi-head self-attention mechanism with simple MLP layers. Additionally, to fully leverage conditional embeddings for guiding the generation process, we regress the scaling parameter $\alpha$ before each residual connection in the sublayers, as shown in the lower left of Figure \ref{pipeline}.

\subsection{Training strategy: Both Generation and Discrimination Perspectives}\label{Train}
The training process of DiffusionCom is divided into two stages. The first stage focuses on training the Structure-MKGformer component. After completing the training of Structure-MKGformer, the second stage involves training for Conditional Denoising. The training of Structure-MKGformer follows the same strategies and parameter settings as MKGformer \cite{MKGformer}, with the primary objective being the optimization of the binary cross-entropy loss function, which is defined as follows:
\begin{equation}
\mathcal{L}_{D}=-[y \cdot \log (p)+(1-y) \cdot \log (1-p)]
\end{equation}
Here, $p$ represents the probability value obtained after applying the Sigmoid function, and $y$ denotes the ground truth label. During the training process of Conditional Denoising, in addition to employing the binary cross-entropy loss, we also introduce the KL divergence loss. This enables joint optimization of Conditional Denoising from both the generative and discriminative perspectives, as detailed below:
\begin{equation}
\mathcal{L}_{G}=\mathbb{E}\left[\operatorname{KL}\left(x_{0} \| f_{\phi}\left(\hat{x}_{K},\hat{x}^{c}\right)\right)\right]
\end{equation}
\begin{equation}
\mathcal{L}_{D}=-[y \cdot \log (Sigmoid(\hat{x})+(1-y) \cdot \log (1-Sigmoid(\hat{x})]
\end{equation}
\begin{equation}
\mathcal{L}_{CD}=\mathcal{L}_{G}+\mathcal{L}_{D}
\end{equation}
Here, $\mathcal{L}_{CD}$ represents the final optimization loss, which comprises the generation loss $\mathcal{L}_{G}$ and the discrimination loss $\mathcal{L}_{D}$.
The training algorithms of DiffusionCom are presented in Algorithm \ref{The first stage} and Algorithm \ref{The second stage}.

\begin{table}[H]
    \caption{Statistics of Dataset.}
        \label{Statistics of Dataset}
    \centering
    \resizebox{0.4\textwidth}{!}{
        \begin{tabular}{cccccc}
            \toprule
            \textbf{Dataset}  & \textbf{\#Rel.} & \textbf{\#Ent.} & \textbf{\#Train}  & \textbf{\#Dev} & \textbf{\#Test}\\
            \midrule
            FB15k-237-IMG      & 237        & 14,541           & 272,115 
            & 17,535          &20,466
            \\    
            WN18-IMG      & 18        & 40,943           & 141,442
            & 5,000         &5,000
            \\    
            \bottomrule
        \end{tabular}
    }
\end{table}
\begin{table*}[ht]
    \centering
    \caption{Performance of Models on FB15k-237-IMG and WN18-IMG datasets. \textbf{Bold} and \underline{underline} highlight the best and second-best performance.}
    \label{tab:my-table}
    \begin{tabular}{cccccccccc}
\hline
\multirow{2}{*}{Model Name}  & \multicolumn{4}{c}{FB15k-237-IMG} &  & \multicolumn{4}{c}{WN18-IMG}    \\ \cline{2-5} \cline{7-10} 
                                                   & MR   & Hits@1  & Hits@3 & Hits@10 &  & MR  & Hits@1 & Hits@3 & Hits@10 \\ \hline
\multicolumn{10}{c}{\textbf{Non-Transformer-based MKGC Models}}                                                               \\ \hline
IKRL(IJCAI 2017)                                     & 298  & 0.194   & 0.284  & 0.458   &  & 596 & 0.127  & 0.796  & 0.928   \\
TransAE(ESWC 2018)                                   & 431  & 0.199   & 0.317  & 0.463   &  & 352 & 0.323  & 0.835  & 0.934   \\
RSME(ACMMM 2021)                                     & 417  & 0.242   & 0.344  & 0.467   &  & 223 & 0.943  & 0.951  & 0.957   \\
MoSE(EMNLP 2022)                                     & \underline{127}  & 0.268   & 0.394  & 0.540   &  & \textbf{7}   & \underline{0.948}  & 0.962  & 0.974   \\
LAFA(AAAI 2024)                                      & \textbf{136}  & \underline{0.269}   & \underline{0.398}  & \underline{0.551}   &  & 25  & 0.947  & 0.965  & 0.977   \\ \hline
\multicolumn{10}{c}{\textbf{Transformer-based MKGC Models}}                                                                   \\ \hline
KG-BERT(2019)                                       & 153  & -       & -      & 0.420   &  & 58  & 0.117  & 0.689  & 0.926   \\
VisualBERT(2019)                                    & 592  & 0.217   & 0.324  & 0.439   &  & 122 & 0.179  & 0.437  & 0.654   \\
VILBERT(ICLR 2020)                                   & 483  & 0.233   & 0.335  & 0.457   &  & 131 & 0.223  & 0.552  & 0.761   \\
MKGformer(SIGIR 2022)                                & 221  & 0.256   & 0.367  & 0.504   &  & 28  & 0.944  & 0.961  & 0.972   \\
SGMPT$^{*}$(ACMMM 2024)                                    & 238  & 0.252   & 0.370  & 0.510   &  & 29  & 0.943  & \underline{0.966}  & \underline{0.978}   \\
MyGO$^{*}$(AAAI 2025)                                    & -    & 0.19    & 0.289  & 0.447   &  & –   & 0.706  & 0.937  & 0.941   \\
MPIKGC$^{*}$(COLING 2024)                            & -    & 0.244   & 0.358  & 0.503   &  & –   & -      & -      & -       \\
AdaMF-MAT$^{*}$(COLING 2024)                         & –    & 0.231   & 0.350  & 0.491  &  & –   & 0.736  & 0.943  & 0.958   \\ \hline
\rowcolor[RGB]{220, 240, 255} DiffusionCom(Ours)                  & 173    & \textbf{0.372}       & \textbf{0.493}      & \textbf{0.617}       &  & \underline{17}  & \textbf{0.973} & \textbf{0.981} & \textbf{0.985}  \\ \hline
\footnotesize{${*}$ indicates the result comes from SGMPT \cite{SGMPT}.}
\end{tabular}
\end{table*}

\section{EXPERIMENTS}
\subsection{Experimental Setups}
\subsubsection{Datasets}
We select two public available multimodal knowledge graph datasets to evaluate the performance of our model: FB15k-237-IMG \cite{WN18} and WN18-IMG \cite{WN18}. Both datasets include three modalities: knowledge graph structures, textual descriptions of entities, and images associated with entities. FB15k-237-IMG is an extended version of the FB15k-237 \cite{FB15K-237} dataset, providing 10 images for each entity. WN18-IMG is an extended version of the WN18 \cite{WN18} dataset, which originates from WordNet \cite{wordnet}, with 10 images added for each entity. The detailed statistics of these datasets are listed in Table \ref{Statistics of Dataset}.

\subsubsection{Implementation Details}
Our model is trained on a single Nvidia A100 GPU, with the code implemented using PyTorch. In the first stage of training, our hyperparameter settings align with those of MKGformer \cite{MKGformer}, with the only difference being our use of a cosine learning rate scheduler instead of a linear one. The cosine learning rate scheduler has been widely adopted in the fine-tuning of many large-scale pre-trained models \cite{roziere2023code,bai2024longalign}. In the second stage, for the FB15k-237-IMG dataset, we set the learning rate to 2e-5, the batch size to 96, the number of diffusion steps to 40, the number of CDenoiser blocks to 1, and the hidden size of the MLP layer in the CDenoiser block to 2048. For the WN18-IMG dataset, the learning rate is set to 3e-5, the batch size to 128, the number of diffusion steps to 30, the number of CDenoiser blocks to 1, and the hidden size of the MLP layer in the CDenoiser block to 1024. During the inference phase, we follow the relevant configurations from \cite{long2024fact} for the experiments.
\begin{figure}[t]
\centering
\includegraphics[width=0.47\textwidth]{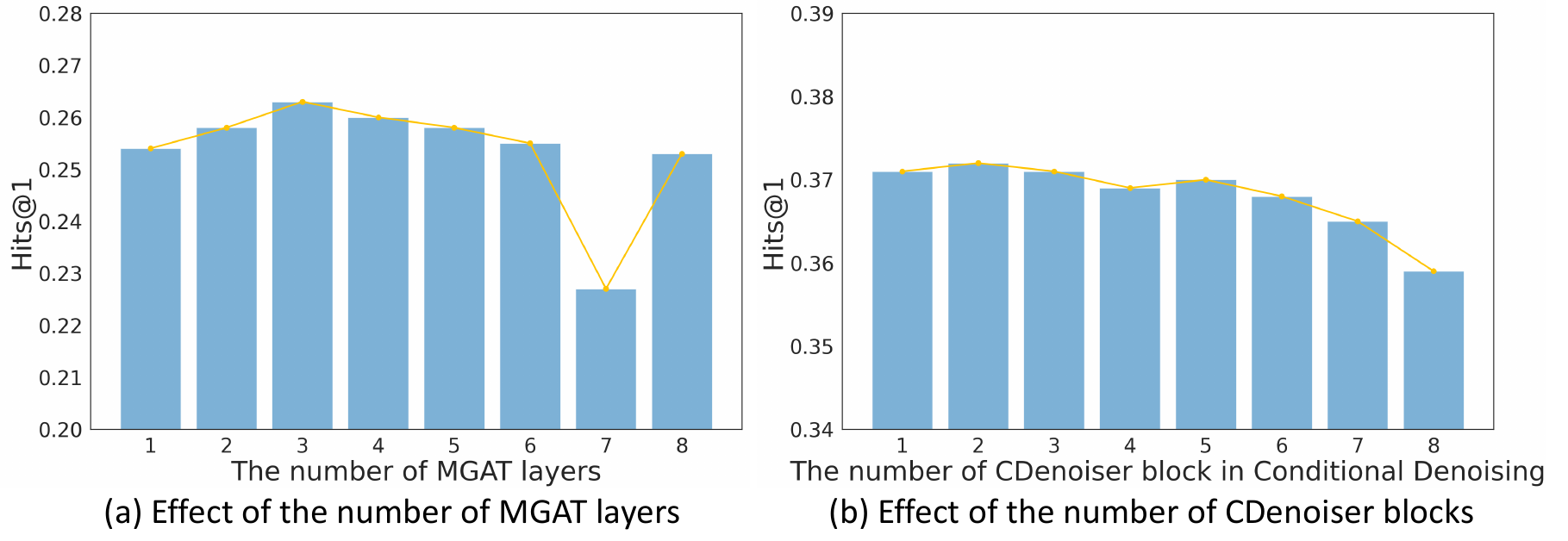}
\caption{Parameter sensitivity analysis on the FB15k-237-IMG dataset with respect to (a) the number of Multimodal Graph Attention Network layers and (b) the number of CDenoiser blocks in Conditional Denoising.}
\label{can1}
\end{figure} 
\subsubsection{Baselines}
To thoroughly evaluate the performance of multimodal knowledge graph completion methods, we select models from two primary categories: non-transformer-based and transformer-based models. The non-transformer-based models include IKRL \cite{IKRL}, TransAE \cite{TransAE}, RSME \cite{RSME}, MoSE \cite{MoSE}, and LAFA \cite{LAFA}. The transformer-based models include KG-BERT \cite{kg-bert}, VisualBERT \cite{visualbert}, VILBERT \cite{Vilbert}, MKGformer \cite{MKGformer}, SGMPT \cite{SGMPT}, MyGO \cite{MyGO}, MPIKGC \cite{MPIKGC}, and AdaMF-MAT \cite{AdaMF-MAT}.

\subsubsection{Evaluation Metrics}
To evaluate the performance of our model on multimodal knowledge graph completion task, we use standard metrics such as Hits@k (including Hits@1, Hits@3, and Hits@10) to measure ranking accuracy at various thresholds, and Mean Rank (MR) to assess the average rank of true entities in the predicted results.

\begin{figure*}[t]
\centering
\includegraphics[width=0.98\textwidth]{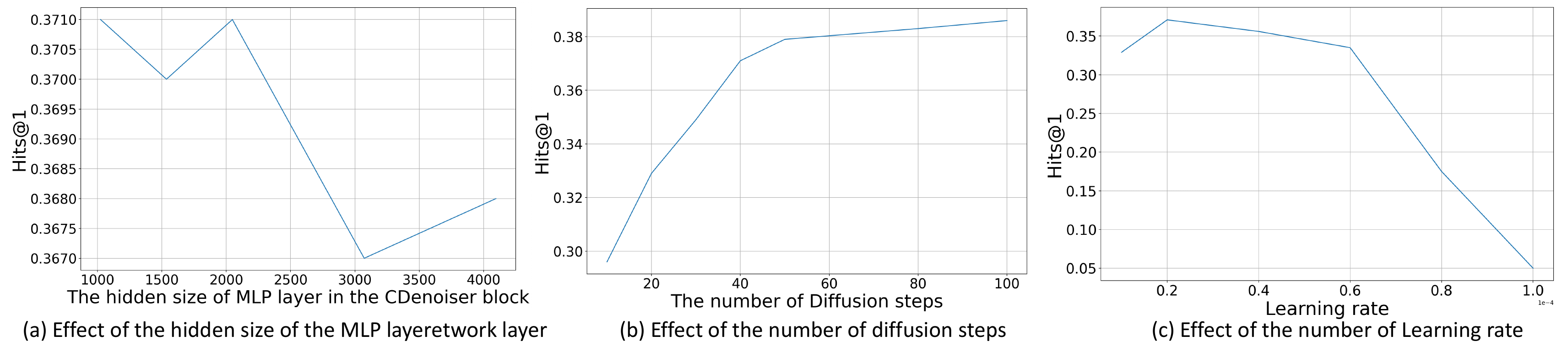}
\caption{Parameter sensitivity analysis on the FB15k-237-IMG dataset with respect to (a) the hidden size of the MLP layer in the CDenoiser block, (b) the number of diffusion steps, and (c) the learning rate.}
\label{can2}
\end{figure*} 

\subsection{Comparison with State-of-the-art}

We compare DiffusionCom with 13 state-of-the-art models across two benchmark datasets. As shown in Table \ref{tab:my-table}, DiffusionCom outperforms all other models across all evaluation metrics, with the exception of the MR metric. In comparison to the current state-of-the-art MKGC models, our model demonstrates significant improvements on the FB15k-237-IMG dataset, with relative gains of 38.2\%, 23.9\%, and 11.9\% in Hits@1, Hits@3, and Hits@10, respectively. Similarly, on the WN18RR-IMG dataset, our model also shows advantages, achieving increases of 2.6\%, 1.6\%, and 0.7\% in Hits@1, Hits@3, and Hits@10, respectively. Although the performance of our model on the MR metric is not the best, it still ranks second. Moreover, compared to the backbone model MKGformer, DiffusionCom demonstrates even better performance on the MR metric. In conclusion, the results show that DiffusionCom excels in multimodal reasoning, confirming the effectiveness of modeling multimodal completion as a progressive generation of joint distributions from noise.

\subsection{Ablation Study}
To validate the effectiveness of each learning component, we conduct a comparative analysis of five DiffusionCom variants on the FB15k-237-IMG dataset. DiffusionCom-MGAT: this variant removes the MGAT module, retaining only the MKGformer as the encoder; DiffusionCom-CDenoiser: in this variant, the CDenoiser Block is removed, and the conditional embedding is directly connected instead; DiffusionCom-C: this variant removes the conditional guidance in the denoising process; DiffusionCom-BCE: this variant removes the binary cross-entropy loss function; DiffusionCom-KL: this variant eliminates the KL divergence loss function.

As shown in Table \ref{compare two commongeneration losses}, DiffusionCom demonstrates superior performance, confirming that the integration of the five learning components significantly enhances the completion capabilities of multimodal knowledge graphs. Specifically, the MGAT incorporates structural information through adaptive weighting, allowing the encoded data to represent not only semantic content but also rich structural features. The CDenoiser Block, introduces conditional information in a structured manner, guiding the diffusion model during the denoising process. The binary cross-entropy loss provides discriminative optimization for the model, while the KL loss further improves overall performance from a generative perspective, enabling an effective fusion of generative and discriminative processes.

In addition, we perform ablation experiments on the Structure-MKGformer model. Specifically, Structure-MKGformer-M refers to the variant without the graph attention mechanism, while Structure-MKGformer-R represents the version that omits subgraph extraction and directly extracts structural information from the entire graph. As shown in Table \ref{The first stage of the ablation Study}, the inclusion of MGAT consistently improves performance across all metrics, underscoring the critical role of structural information in multimodal knowledge graph completion. Moreover, extracting structural information from subgraphs is also demonstrated to be essential.

\begin{table}[t]
\centering
\caption{Performance of different DiffusionCom variants on the FB15k-237-IMG dataset.}
\label{compare two commongeneration losses}
\begin{tabular}{cccccc}
\hline
Model               &   MR  & Hits@1 & Hits@3 & Hits@10  \\ \hline
DiffusionCom        & \textbf{173} & \textbf{0.372}  & \textbf{0.493}  & \textbf{0.617}       \\ 
DiffusionCom-MGAT   & 193 & 0.353  & 0.480  & 0.604        \\ 
DiffusionCom-CDenoiser    & 208 & 0.319  & 0.451  & 0.558        \\ 
DiffusionCom-C      & 227 & 0.275  & 0.387  & 0.521      \\
DiffusionCom-BCE      & 252 & 0.244  & 0.352  & 0.485       \\ 
DiffusionCom-KL     & 185 & 0.346  & 0.468  & 0.593       \\  
\hline
\end{tabular}
\end{table}

\begin{table}[t]
\caption{Performance of different Structure-MKGformer variants on the FB15k-237-IMG dataset.}
\label{The first stage of the ablation Study}
\begin{tabular}{cccccc}
\hline
Model       &  MR         & Hits@1         & Hits@3 & Hits@10  \\ \hline
Structure-MKGformer    & \textbf{218} & \textbf{0.262}  & \textbf{0.379}     & \textbf{0.512}     \\ 
Structure-MKGformer-M  & 280  & 0.253       & 0.363  & 0.502         \\
Structure-MKGformer-R  & 225  & 0.254       & 0.367 & 0.506 \\
\hline
\end{tabular}
\end{table}

\begin{figure*}[t]
\centering
\includegraphics[width=1\textwidth]{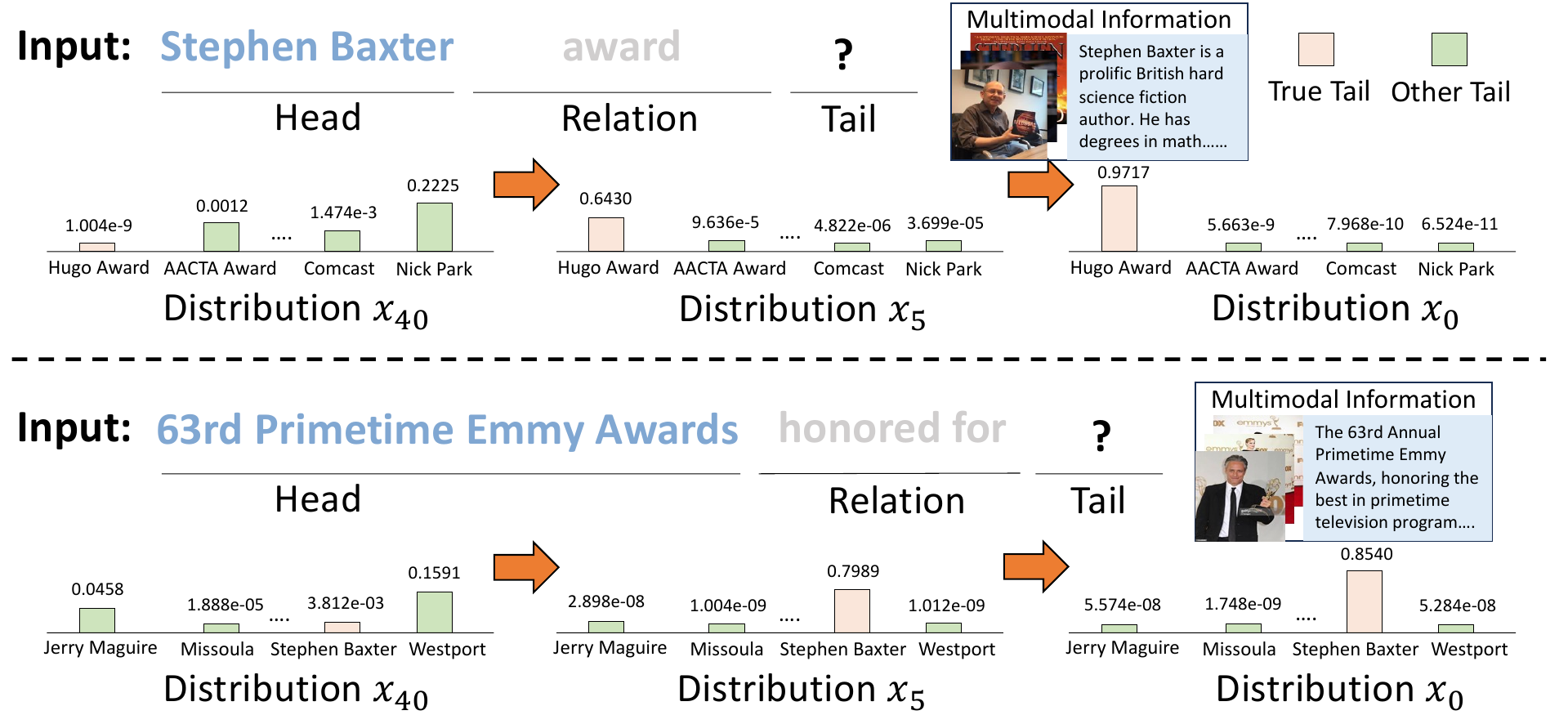}
\caption{The visualization of the probability distribution diffusion process. We use light pink to indicate the true tail entities and provide a detailed illustration of the step-by-step evolution, from the randomly initialized noise input $x_{40}$ to the final predicted distribution $x_0$.}
\label{tu4}
\end{figure*} 

\subsection{Parameter Sensitivity Analysis}
\subsubsection{The Number of Multimodal Graph Attention Network layers}
In Figure \ref{can1}(a), we explore the impact of the number of Multimodal Graph Attention Network (MGAT) layers on the performance of the Structure-MKGformer model. Specifically, we experimented with MGAT architectures ranging from 1 to 8 layers. The experimental results show that as the number of layers increases, the model performance initially improves but then declines, with the best performance observed at 3 layers. Therefore, we ultimately chose a 3-layer Graph Attention Network as the structural information encoder to achieve optimal encoding results.
\subsubsection{The Number of CDenoiser Block in Conditional Denoising}
In Figure \ref{can1}(b), we analyze the impact of the number of CDenoiser blocks on the model performance. The results indicate that the number of CDenoiser block does not have a significant impact on the performance. Moreover, as the number of blocks increases, the model performance does not improve but instead shows a downward trend. 
\subsubsection{The Hidden Size of MLP Layer in the CDenoiser Block}
In Figure \ref{can2}(a), we analyze the impact of the hidden size of the MLP layers in the CDenoiser module on the model performance. In the experiments, it is found that appropriately increasing the size of the hidden units helps to improve the model performance. However, when the scale of the hidden units becomes excessively large (e.g., 2048), the model performance decreases instead.
\subsubsection{The Number of Diffusion Steps}
In Figure \ref{can2}(b), we analyze the impact of the number of diffusion steps on task performance. The experimental results indicate that in the task of multimodal knowledge graph completion, performance improvements become extremely limited after more than 50 diffusion steps. This contrasts significantly with the 1,000 diffusion steps commonly used in image generation tasks\cite{dhariwal2021diffusion,wang2022zero}. We hypothesize that this phenomenon may be due to the relatively simple probability distribution in the completion task, which differs from the complex pixel distributions in natural images. Therefore, fewer diffusion steps are required to achieve optimal results in the multimodal knowledge graph completion task compared to image generation tasks.
\subsubsection{Learning Rate}
In Figure \ref{can2}(c), we conduct a detailed search for the learning rate within the range of $[1e-5, 2e-5, 4e-5, 6e-5, 8e-5, 1e-4]$. The experimental results show that the model performs best when the learning rate is set to 2e-5. However, when the learning rate exceeds 8e-5, the model performance drops sharply, and at a learning rate of 1e-4, the model training even fails to complete successfully.

\subsection{Advantages of Diffusion Models in MKGC}
Diffusion models have demonstrated exceptional generative capabilities across various domains. Beyond their impressive generative performance, we further explore their unique advantages in the task of multimodal knowledge graph completion, where they stand out compared to other generative methods. The core strength of diffusion models lies in their stepwise generation process, progressively refining the relation between the triple $(h, r, ?)$ and candidate tail entity $t$, from a coarse to a fine level. This gradual approach enables diffusion models to be more efficient in this task compared to generative models such as Generative Adversarial Networks (GANs) \cite{goodfellow2020generative} and Variational Autoencoders (VAEs) \cite{doersch2016tutorial}. To better understand the diffusion process, we present a visualization of the diffusion process in Figure \ref{tu4}. With the reverse diffusion process, the noise gradually generates a credible joint probability distribution and finally successfully predicts. These results demonstrate that our method can progressively reveal the correlations, further validating its significant advantages in the multimodal knowledge graph completion task.
\section{CONCLUSION}
In this paper, we propose DiffusionCom, a novel framework that represents the first diffusion model-based approach for multimodal knowledge graph completion (MKGC) task. It is also the first to tackle MKGC from a generative perspective. DiffusionCom explicitly models the joint probability distribution between the (head, relation) pair and candidate tail entities, overcoming the inherent limitations of existing discriminative methods when dealing with complex multimodal knowledge graphs. To effectively extract structural information from multimodal knowledge graphs and integrate it into DiffusionCom, we propose a structure-aware multimodal knowledge representation learning method based on a multimodal graph attention network, called Structure-MKGformer. This method enables reasoning over multimodal knowledge graphs, deeply mining and adaptively fusing fine-grained latent structural relationships between entities, thereby further highlighting the critical role of structural information in knowledge graph completion task. Moreover, we optimize DiffusionCom from both generative and discriminative perspectives, providing it with dual advantages. Extensive experiments show that DiffusionCom significantly outperforms all current mainstream methods on two widely-used datasets: FB15k-237-IMG and WN18-IMG.

\bibliographystyle{ACM-Reference-Format}
\bibliography{sample-base}
\end{document}